\documentclass[twocolumn,pra,showpacs,floatfix]{revtex4}
\usepackage{amsmath}
\usepackage{amsfonts}
\usepackage{amsthm}
\usepackage{graphicx}
\usepackage{amssymb}

\newcommand{\degree}[1]{\ensuremath{#1^\circ}}

\begin{document}
\title{Single qubit gates with Jump and Return sequences}
\author{M.~D.~Bowdrey}\email{mark.bowdrey@physics.org}
\affiliation{Centre for Quantum Computation, Clarendon Laboratory,
University of Oxford, Parks Road, OX1 3PU, United Kingdom}
\author{J.~A.~Jones}\email{jonathan.jones@qubit.org}
\affiliation{Centre for Quantum Computation, Clarendon Laboratory,
University of Oxford, Parks Road, OX1 3PU, United Kingdom}

\date{\today}
\pacs{03.67.Lx,82.56.-b}
\begin{abstract} We discuss the
implementation of frequency selective rotations using sequences of hard pulses and delays. These rotations are suitable for implementing single qubit gates in Nuclear
Magnetic Resonance (NMR) quantum computers, but can also be used in other related implementations of quantum computing. We also derive methods for implementing hard
pulses in the presence of moderate off-resonance effects, and describe a simple procedure for implementing a hard $180^\circ$ rotation in a two spin system. Finally we
show how these two approaches can be combined to produce more accurate frequency selective rotations.
\end{abstract}
\maketitle

\section{Introduction}
A key element in any implementation of quantum computing \cite{bennett00} is the ability to perform single qubit gates. Clearly these require the ability to rotate one
qubit (spin) while leaving the rest of the system unaffected.  In many proposed implementations this is achieved using the fact that qubits are spatially localised, but
in liquid state Nuclear Magnetic Resonance (NMR) systems \cite{ernst87} this is not possible, and it is instead necessary to distinguish the qubits by their different
resonance frequencies. Thus frequency selective excitation plays a key role in NMR quantum information processing \cite{cory96, cory97, gershenfeld97, jones98c,
jones01a, vandersypen04}.

The simplest form of excitation in an NMR experiment is a short high power burst of radiofrequency radiation, known as a ``hard'' rf pulse. Such a pulse will excite
spins over a considerable range of frequencies: a simple (if not entirely correct) argument \cite{ernst87} is that the frequency domain excitation profile is the
Fourier transform of the time domain pulse shape.  Frequency selective excitation could in principle be achieved by using a long low power pulse (a ``soft'' pulse), but
the selectivity is poor as the Fourier transform of a rectangular pulse is a sinc function, which has considerable power in the wings.  A better method is to use shaped
pulses, instead of constant power rectangular pulses, thus producing a more desirable profile in the frequency domain \cite{freeman97b}.

This approach has been extensively developed, both for conventional NMR experiments \cite{freeman97b} and for NMR quantum information processing
\cite{vandersypen01,fortunato02} (note that the properties of many modern shaped pulses cannot be rationalized using Fourier arguments and it is frequently necessary to
use numerical methods to calculate them). In some experiments, however, it is overcomplex. Consider the case where a hard excitation pulse will only affect two qubits,
for example in an NMR system which only has two spins of a particular nuclear species. In this case a much simpler scheme based on hard pulses and delays can be applied
\cite{jones99,cummins02,anwar04}.

\section{Jump and Return}
Schemes for frequency selective rotation using hard pulses are ultimately based on the ``Jump and Return'' solvent suppression pulse sequence \cite{gueron82,hore83},
which is designed to avoid exciting signals directly resonant with the applied radiation while (at least partially) exciting all other spins. To see how the Jump and
Return sequence can be modified for selective rotation, it is helpful to show how this sequence works for solvent suppression, which is effectively selective
non-rotation.

\begin{figure}
\begin{center}
\includegraphics{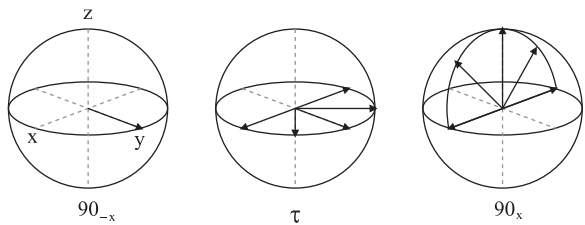}
\caption{Jump and Return solvent suppression sequence: the solvent signal (at zero frequency in the rotating frame) experiences no overall evolution, while spins at
other frequencies are partially excited.} \label{SolventSuppression}
\end{center}
\end{figure}

The Jump and Return sequence comprises a period of free precession sandwiched by two $90^\circ$ hard pulses on resonance with the solvent frequency,
\begin{equation}
90_{-x} \; \tau \; 90_{x}. \label{tauJR}
\end{equation}
This and subsequent pulse sequences are written in NMR shorthand notation, where $90$ is the nutation angle in degrees and subscripts indicate the nutation axis, which
for ideal pulses lies in the $xy$-plane; note that time runs from left to right, so that the $90_{-x}$ pulse is applied first. We assume that the system begins in its
thermal equilibrium state, with all the spins along the $z$-axis of the Bloch sphere \cite{freeman97b}. The first hard pulse excites all the spins, and during the
period of free precession any spins away from the solvent frequency will precess. The solvent spins do not precess in the rotating frame \cite{ernst87, freeman97b}, and
are returned to the initial state by the second pulse, but non-solvent spins that have strayed during the period of free precession are not fully returned, and hence
result in some excitation (see figure \ref{SolventSuppression}).  It is easily shown that the sequence in Eq.~\ref{tauJR} is equivalent to a $\theta_{-y}$ hard pulse,
where $\theta$ is the precession angle during the period $\tau$.

The amount of excitation that a spin receives thus depends on the offset of the spin frequency from the pulse frequency, as for a given duration between pulses, the
extent of the offset corresponds to an angle of rotation in the $xy$-plane.  A spin will be maximally excited if it has precessed $90^\circ$ from its initial position
in the $xy$-plane (and therefore is completely unaffected by the second pulse), as shown in figure \ref{SolventSuppression}.  There are two offsets where spins will
precess $90^\circ$ in the $xy$-plane (\emph{i.e.}, be maximally excited) and these are equally distant above and below the pulse frequency.  Spins at these positive and
negative offsets will end up along the same axis in the $xy$-plane but pointing in opposite directions, corresponding to overall positive and negative rotations
respectively.

Now we can return to our original situation with only two spins and consider how the Jump and Return sequence can be useful for selective rotation.  The reference
frequency is set halfway between the resonance frequencies of the two spins, so that in the rotating frame the spins will precess at equal but opposite rates,
$\pm\delta\omega$.  The Hamiltonian can then be written using product operator notation \cite{sorensen83} as
\begin{equation}
{\mathcal{H}}=\delta\omega(I_z-S_z)+\pi J\, 2I_zS_z
\end{equation}
where $J$ is the spin--spin coupling and we have assumed a weakly coupled system, where $|\delta \omega| >> |J|$.  The effect of weak coupling is to replace each
transition frequency by a cluster of transitions, known as a multiplet, but if the width of the multiplet is narrow compared with the separation between multiplets
(which will be true for weak coupling) then the effects of spin--spin coupling can be neglected.  We also assume for the moment that the hard pulses are instantaneous,
so that there is no precession during a pulse (we will return to this assumption later). If we set the delay period between the pulses to $\tau=\pi/2\delta\omega$ the
two spins will rotate by $\pm90^\circ$ during the precession period, corresponding to the offset resulting in maximum excitation in the solvent suppression case.

So far we have largely considered excitation sequences, that is sequences designed to take a spin from the $z$-axis into the $xy$-plane, but these sequences can be used
to perform the general rotations required for quantum computing, as the Jump and Return sequence can be viewed as a simple rearrangement of the composite $z$-pulses
(sequences designed to perform $z$-rotations using only $xy$-rotations) devised by Freeman \emph{et al}. \cite{freeman81}. In general, one can perform any rotation
around the Bloch sphere using rotations that are orthogonal to the desired rotation.  In particular
\begin{equation}
\label{JR} 90_{-y} \;  \theta_{[+z,-z]} \; 90_{y} \; \equiv \;
[\theta_{x}\, , \, -\theta_{x}]
\end{equation}
where we use non-selective $y$-rotations (hard pulses), to convert a contra-axial $z$-rotation (an appropriate period of free precession) into a contra-axial
$x$-rotation.  Rotations with other phase angles can, of course, be achieved by changing the phases of the two outer pulses. For a selective, rather than contra-axial,
$xy$-rotation we can apply another non-selective $xy$-rotation along the same axis as the contra-axial $xy$-rotation \cite{jones99}, so that
\begin{equation}
\label{SelJRExtraPulse} 90_{-y} \; \frac{\theta}{2}_{[+z,-z]} \;
90_{y} \; \frac{\theta}{2}_{x} \; \equiv \; [\theta_{x}\, , \,
\openone]
\end{equation}
where $\openone$ indicates the identity (do-nothing) operation. The extra non-selective rotation will return the spin we do not wish to rotate to its initial position,
and further rotate the spin we want to select. Note that we halve the initial contra-axial rotation angle to compensate for the extra rotation on the selected spin.

There is a small but useful improvement to the previous sequence: instead of adding a \textit{non-selective} hard pulse to convert the contra-axial $xy$-rotation into a
selective $xy$-rotation, we can apply a \emph{non-selective} $z$-rotation, that acts equally on both spins, to reset one spin,
\begin{equation}
\label{OppAndNonSelZrot} 90_{-y} \; \frac{\theta}{2}_{[+z,-z]} \;
\frac{\theta}{2}_{+z} \; 90_{y} \; \equiv \; [\theta_x\, , \,
\openone]
\end{equation}
For the spin that we wish to return to its original state, the non-selective $z$-rotation cancels the $z$-rotation due to precession.  Thus this spin does not, in
effect, precess between hard pulses and so is returned by the second hard pulse.  Note that it is not actually necessary to perform the non-selective $z$-rotation;
instead we can simply move into a new reference frame, effectively rotating the frame rather than the spins \cite{knill00}, thus saving one pulse. Equivalently, we can
pass the $z$ rotation through the pulse sequence, so that the selective Jump and Return sequence (equation \ref{OppAndNonSelZrot}) becomes
\begin{equation}
\label{JRselectiveRotation} 90_{\phi-90} \;
\frac{\theta}{2}_{[+z,-z]} \; 90_{\phi+90-\frac{\theta}{2}} \;
\frac{\theta}{2}_{+z} \; \equiv \; [\theta_{\phi}\, , \, \openone]
\end{equation}
where subscripts on pulses now indicate phase angles in the $xy$-plane.  Each $z$ rotation can be passed through successive pulses until it reaches the very end of the
sequence, at which point the final rotation can either be implemented using a composite $z$-pulse \cite{freeman81} or, if the calculation ends with qubits in an
eigenstate, simply be dropped.  In some cases it is preferable to develop the pulse sequence as a series of independent modules, and in this case $z$ rotations should
be moved to the end of each module and then corrected using composite pulses.

These techniques can be extended to provide a wide repertoire of single qubit gates.  By the combination of contra-axial $z$-rotations (free precession) and
non-selective $z$-rotations (frame rotations) we can perform selective $z$-rotations on either spin.  In this way we can selectively rotate one spin about the $z$-axis
to perform equally long $xy$-rotations on both spins, but around different $xy$ axes for each spin,
\begin{multline}
\label{JRdifferentPhase} \frac{\alpha}{2}_{-z}
\frac{\alpha}{2}_{[+z,-z]}  \theta_{\phi}
\left(2\pi-\frac{\alpha}{2}\right)_{-z}
\left(2\pi-\frac{\alpha}{2}\right)_{[+z,-z]} \\
\equiv [\theta_\phi  , \theta_{\phi + \alpha}].
\end{multline}
Using methods of this kind any desired combination of rotation and phase angles can be implemented.  Alternatively, such combinations can be constructed by using two
selective pulses in succession.

\section{Solvent suppression sequences}
The Jump and Return sequence is the simplest member of a large family of solvent suppression sequences \cite{hore83} developed for use in conventional NMR. Although
Jump and Return does not excite signals exactly on resonance, this selectivity is not very robust, and signals only a small distance from resonance \textit{are}
significantly excited.  A family of sequences, known as binomial solvent suppression sequences \cite{hore83}, have been developed which have much broader nulls in their
excitation profiles.

The best know of these sequence is usually called $1\bar{3}3\bar{1}$ and for a $90^\circ_x$ rotation has the form
\begin{equation}
\alpha_x \; 2\tau \; (3\alpha)_{-x} \; 2\tau \; (3\alpha)_x \; 2\tau
\; \alpha_{-x} \label{tau1331}
\end{equation}
where the pulse lengths are chosen such that $8\alpha=\degree{90}$, and $\tau=\pi/2\delta\omega$ as before; the bars over pulse lengths in the name correspond to the
changes in the pulse phases. Although this sequence is in some ways related to Jump and Return its basic operation differs in an important way: in Jump and Return
sequences the target spin is fully excited by the first \degree{90} pulse, while in binomial sequences the excitation builds up steadily throughout the pulse sequence.
This leads to the use of a precession time $2\tau$, compared with $\tau$ in Jump and Return.

As binomial sequences give more robust solvent suppression than Jump and Return, it might seem that they would also give more robust selective pulses, showing less
dependence on small deviations from ideality in their implementation.  There are, however, three reasons to doubt this surmise.  Firstly, the binomial family is
designed to give robust non-excitation of the solvent signal, not robust excitation of the target spin.  Secondly, the binomial family is designed solely for the
purpose of excitation, and it is not clear how well it will behave as a general rotor \cite{cummins01b}. Finally the much longer evolution periods involved ($6\tau$ for
$1\bar{3}3\bar{1}$, compared with $\tau$ for a simple Jump and return sequence), mean that these sequences are likely to be vulnerable to errors in the value of
$\delta\omega$, such as those arising from multiplet structure.

\begin{figure}
\begin{center}
\includegraphics[scale=0.5]{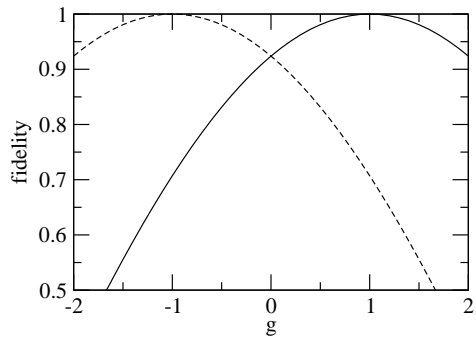}
\caption{Propagator fidelity for a selective $90^\circ$ pulse implemented using the Jump and Return approach of Eq.~\ref{SelJRExtraPulse} as a function of the ratio $g$
between the actual offset frequency and its nominal value.  The solid line shows the fidelity with respect to a $90^\circ$ pulse, while the dashed line shows the
fidelity against an identity operation.  These curves are given by $\cos[\pi(1\mp g)/8]$ respectively.} \label{fidelity}
\end{center}
\end{figure}

We have investigated this question by modifying a range of binomial solvent suppression sequences to give selective excitation sequences, and then calculating
propagator fidelities \cite{bowdrey02} as a function of small errors in $\delta\omega$. Although the detailed results are quite complex \cite{bowdrey03}, the overall
result is simple: selective excitations based on Jump and Return sequences are far more robust than those based on more sophisticated solvent suppression sequences. For
this reason we will only consider Jump and Return sequences for the remainder of this paper.  A fidelity plot for one particular Jump and Return sequence is shown in
Fig.~\ref{fidelity}.

\section{Off-resonance effects} \label{Sec:Off-Resonance}
\begin{figure}
\begin{center}
\includegraphics[scale=0.8]{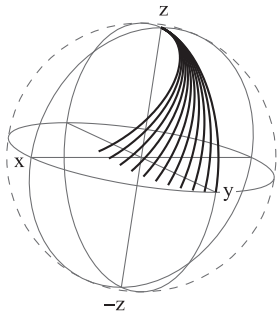}
\caption{Off-resonance effects on a $90^\circ$ excitation pulse for off-resonance fractions $f$ in the range 0 to 1 in steps of $0.1$.} \label{OffRes90}
\end{center}
\end{figure}

Although Jump and Return based sequences are generally quite successful when implementing single qubit gates, there is a fundamental problem in their design.  Since we
use the difference in resonance frequencies to generate contra-axial rotations, it is obviously impossible for our hard pulses to be perfectly on-resonance with both
transitions.  In general we can parameterise off-resonance effects by the off-resonance fraction \cite{cummins03}, $f=\delta\omega/\omega_1$, where $\delta\omega$ is
the off-resonance frequency, and $\omega_1$ is the nutation rate induced by the RF pulse.  The corresponding propagator for an $x$ rotation is
\begin{equation}\begin{split}
U&=\exp[-i\omega_1t(I_x+fI_z)]\\
 &=\begin{bmatrix}
 \cos(\zeta)-if\sin(\zeta)/\gamma&-i\sin(\zeta)/\gamma\\
-i\sin(\zeta)/\gamma&\cos(\zeta)+if\sin(\zeta)/\gamma
 \end{bmatrix}
\end{split}\label{OffResProp}\end{equation}
where $\zeta=\gamma\omega_1t/2$ and $\gamma=\sqrt{1+f^2}$.  When $f=0$ this reduces to the simple form
\begin{equation}
\begin{bmatrix}
 \cos(\omega_1t/2)&-i\sin(\omega_1t/2)\\
-i\sin(\omega_1t/2)&\cos(\omega_1t/2)
 \end{bmatrix}
\end{equation}
as expected. When $f\ne0$ a pulse will induce a rotation around an axis tilted away from the $xy$-plane as shown in Fig.~\ref{OffRes90}. However, the undesired effect
of rotating around this tilted axis is a systematic error and can be compensated for by various means, including composite pulses.

The use of composite pulses to correct for off-resonance errors has been widely studied in conventional NMR \cite{levitt86}. More recently, corrective composite pulses
that act as general rotors, known as Class-A composite pulses \cite{levitt86}, have been developed for conventional NMR experiments \cite{wimperis94} and for NMR
quantum computing \cite{cummins03}. For a system with just two spins, with equal and opposite known resonance offsets, there is an analytical solution for a composite
pulse that perfectly corrects the off-resonance error.  This \textsc{rotten} (Resonance Offset Tailoring To Enhance Nutations) composite pulse \cite{cummins01b}
consists of a time symmetric sequence of three hard pulses, that carries out the desired $xy$-rotation on both spins.  The fidelity of the desired rotation at the
actual resonance offset of both spins is unity, since the \textsc{rotten} composite pulse sequence is tailored to perform perfectly for this resonance offset.

Although it is possible to combine the three period \textsc{rotten} sequence with the Jump and Return approach, a simpler sequence results if two of the RF pulses are
converted to $z$-rotations. A helpful insight into how this new sequence works is provided by an existing result from conventional NMR.  It is well known that a
\degree{90} pulse with a small off-resonance fraction $f$ applied  to an initial $z$-state (usually called an excitation pulse) is, to a good approximation, equivalent
to an ideal \degree{90} hard pulse followed by a period of free precession for a time equal to $2/\pi\approx0.64$ of the pulse length \cite{cavanagh}. This can be seen
in Fig.~\ref{OffRes90}, where the lines corresponding to small values of $f$ all terminate near the $xy$-plane, and are offset from the $y$ axis by angles that are
roughly proportional to the off-resonance fraction.  The same conclusion can be reached by expanding Eq.~\ref{OffResProp} to first order in $f$.

This approach is not confined to excitation but can be generalized: a \degree{90} pulse can be approximated as an ideal pulse surrounded by free precession periods of
length equal to $2/\pi$ times the pulse length.  (The case of an excitation pulse, described above, is special: the initial precession period can be dropped as
$z$-states are invariant under free precession, allowing a simpler approximation to be used.) Reversing this, it is clear that a \degree{90} pulse applied to a spin
with a small off-resonance fraction $f$ would behave almost like an ideal pulse if it were preceded and followed by periods of length equal to $-2/\pi$ times the pulse
length; methods for implementing \textit{negative} time periods are considered below.

For larger values of $f$ this approximation begins to break down in two ways.  Firstly the phase angle is no longer linear in the offset frequency, and so it is not
possible to correct this using a fixed evolution time, and secondly the final state can lie some distance from the $xy$-plane.  It is, however, possible to achieve an
ideal rotation for any particular off-resonance fraction in the range $|f|\le1$ by placing negative $z$-rotations before and after a pulse with a nominal nutation angle
greater than \degree{90}.  The $z$-rotation sandwich
\begin{equation}
\phi'_z \; \theta'_x \; \phi'_z
\end{equation}
(where $\theta'$ is the nutation angle of the central pulse in the \textit{absence} of off-resonance effects) can be shown to perform an ideal $90_x$ rotation as long
as the angles are chosen as
\begin{equation}
\label{z90z} \phi'=-\arcsin(f)\qquad
\theta'=\frac{\arccos(-f^2)}{\sqrt{1+f^2}}.
\end{equation}
Pulses with other phase angles can, of course, be achieved by changing the phase of the central pulse.  For small off-resonance fractions Eq.~\ref{z90z} can be expanded
to first order in $f$, giving $\phi'\approx-f$ and $\theta'\approx\pi/2$.  Outside the range $|f|\le1$ the angles $\theta'$ and $\phi'$ become imaginary, and this
approach is no longer possible.

This approach can also be used for two spins with equal and opposite resonance offsets; the nutation angle $\theta'$ will be the same for both spins, while the
$z$-rotation angle $\phi'$ will take equal and opposite values for the two spins; the two spins need a period of free precession given by
\begin{equation}
\tau'=\frac{2\phi'}{\pi f}\times t_{90}
\end{equation}
(where $t_{90}$ is the length of the \degree{90} pulse) to rotate in opposite directions. For small values of $f$, $\tau'=(-2/\pi)t_{90}$ as expected.  The spin
precession time is negative, but we can achieve this by subtracting the desired negative precession time from any neighbouring positive precession period. If no such
period exists then one can be created: the two spins will precess through angles of \degree{\pm360} during a time $4\tau$, and such time periods can be inserted with
impunity. In effect this allows the spin to achieve a negative precession angle by going the long way around the Bloch sphere.  In fact it is not necessary to go all
the way round, as a similar effect can be achieved by combining a precession period of length $2\tau$, which gives rotations of \degree{\pm180}, with a simultaneous
$180_z$ rotation, so that one spin is left alone while the other rotates through \degree{360}.

The strategy of a $z$-rotation sandwich also works for a pulse with an arbitrary nutation angle, $\theta_x$.  The required rotations are now given by
\begin{equation}
\phi''=-\arcsin(f\tan[\theta/2])
\end{equation}
and
\begin{equation}
\theta''=\frac{\arccos(-f^2+[1+f^2]\cos\theta)}{\sqrt{1+f^2}}
\end{equation}
with first order approximations $\phi''\approx-f\tan(\theta/2)$ and $\theta''\approx\theta$.  Solutions exist for off-resonance fractions in the range
$|f|\le\cot(\theta/2)$; note that as $\theta$ increases, the range of off-resonance frequencies which can be tolerated decreases. This is particularly obvious for a
\degree{180} rotation (a \textsc{not} gate), which can only be formed when $f=0$, as any tilt in the rotation axis will make it impossible to rotate a state from the
north pole of the Bloch sphere to the south pole.  At the other extreme it is possible to use $z$-rotation sandwiches to perform accurate \degree{5} rotations, even
with off-resonance fractions greater than 20!

A solution to this problem is to implement pulses with large rotation angles as two or more pulses with smaller rotation angles. For example it is possible to implement
a \degree{180} pulse by combining two \degree{90} pulses, which are themselves implemented using $z$-rotation sandwiches.  Thus
\begin{equation}
\tau'' \; \theta''_x \; (4\tau+2\tau'') \; \theta''_x \; \tau''
\end{equation}
where $\tau''=(2\phi''/\pi f)\times t_{90}$ will implement a $2\theta_x$ pulse for two spins with equal and opposite off-resonance fractions.  The relationship of
$\tau''$ to $\theta''$ and $\phi''$ is analogous to that between $\tau'$, $\theta'$ and $\phi'$

The $z$-rotations at either end of the sequence are backward precession times, but it may be possible to combine these with surrounding precession periods. The range of
$f$ values which can be tolerated will be at least doubled by this process, and for large angles the improvement will be much greater. The process can, of course, be
extended indefinitely, dividing a pulse up into a very large number of smaller pulses; the final result is similar to a \textsc{dante} pulse \cite{freeman97b}, but for
such large off-resonance frequencies it is probably more sensible to use conventional shaped pulses.

For the special case of a \degree{180} pulse (simultaneous \textsc{not} gates on both spins), the outer $z$-rotations can simply be dropped.  In general a $180^\circ$
rotation refocuses equal periods of precession on either side of the pulse, so we can add periods of precession to either side and it remains a $180^\circ$ rotation.
Thus we can ``add'' periods of the right length to exactly cancel the backward $z$-precession periods, and the sequence
\begin{equation}
\theta'_x \; (4\tau+2\tau') \; \theta'_x
\end{equation}
will implement a $180_x$ pulse for two spins, and is valid for the
range $|f|\le1$.

This sequence can be shortened by replacing the central \degree{\pm360} evolution by a \degree{\pm180} evolution and a \degree{180} frame rotation, to give the sequence
\begin{equation}
\theta'_x \; 90_z \; (2\tau+2\tau') \; 90_z \; \theta'_x
\end{equation}
where the \degree{180} frame rotation has been split into two.  For the permitted range of off-resonance frequencies ($|f|\le1$) the evolution period will always be
greater than or equal to zero.  The frame rotations can be passed through the pulses to the beginning and end of the sequence, and can then be dropped. Thus the
sequence
\begin{equation}
\theta'_{-y} \; (2\tau+2\tau') \; \theta'_y \label{eq:hard180}
\end{equation}
will also implement a $180_x$ inversion pulse for two spins with $|f|\le1$.

This approach can also be combined with conventional Jump and Return sequences, to perform more accurate selective or contra-axial rotations, by simply replacing each
\degree{90} pulse in a sequence with a $z$-rotation sandwich.  As usual it is possible to combine the different periods of precession to give a simpler sequence. Thus
using equation~\ref{OppAndNonSelZrot} and choosing $\theta=\degree{90}$ gives
\begin{equation}
\label{shortJRz90z} \tau' \; \theta'_{-y} \; (\tau/2+2\tau') \; 45_z
\; \theta'_y \; \tau' \equiv \; [90_x\, , \, \openone]
\end{equation}
where the central evolution periods have been combined, and the $45_z$ rotation can be implemented using frame rotations.   The precession times at either end of the
sequence are negative times, but as before it will be possible to combine these with surrounding precession periods or, failing this, to introduce additional precession
periods of $4\tau$ with which they can be combined.

Once again it is instructive to consider the special case of a $180^\circ$ contra-axial rotation, when the outer $z$-rotations can simply be dropped as before to get
\begin{equation}
\theta'_{-y}\; (2\tau+2\tau') \; \theta'_y
\end{equation}
a sequence identical to that derived previously, equation~\ref{eq:hard180}.  This makes sense: a $180^\circ$ contra-axial rotation is indistinguishable from a
$180^\circ$ hard rotation, as when going half way round the Bloch sphere it does not matter which direction is taken.

\section{Summary}
Jump and Return sequences provide a simple but effective way of implementing frequency selective pulses in two spin systems. They have already proved useful in NMR
quantum computation, and could also be applied in related systems where implementing conventional shaped pulses is impractical. The simplest approach is only suitable
when the difference in resonance frequencies is small compared with the nutation rate of the external field, but more complex approaches can be used to implement both
selective and non-selective pulses in spin systems with larger off-resonance fractions.

Given the power of existing shaped pulse approaches it is important to consider what advantages are offered by Jump and Return sequences, which can be summarized as
simplicity and time-efficiency.  Although modern NMR spectrometers are usually capable of implementing complex shaped pulses, older equipment has much more limited
capabilities, and Jump and Return pulses have been used to implement NMR quantum computing on such devices \cite{jones99,cummins02}.  More importantly electron spin
resonance (ESR) spectrometers are not yet capable of implementing complex shaped pulses, and so simpler approaches will be needed if current quantum computing
experiments based on ESR \cite{morton05} and electron-nuclear double resonance (ENDOR) \cite{morton06} are to be extended.  Jump and Return sequences based on
equation~\ref{SelJRExtraPulse} are particularly simple, requiring only pulses with simple phase angles and delays. Finally, Jump and Return sequences are also
time-optimal in that they are the shortest possible sequences capable of achieving selective excitation, thus minimising decoherence effects during pulses.

\section*{Acknowledgements}
We thank the UK EPSRC and BBSRC for financial support.

\end{document}